\documentclass[aps,pre,floats,superscriptaddress,showpacs,twocolumn]{revtex4-1}

\usepackage{amsmath, amsthm, amssymb}
\usepackage{enumitem}
\usepackage{graphics,graphicx}
\usepackage{epsfig}
\usepackage{times}
\usepackage{dcolumn}
\usepackage{bm}
\usepackage{url}

\usepackage{ulem, color}

\begin{document}

\title{Optimal phase synchronization in networks of phase-coherent chaotic oscillators}

\author{P. S. Skardal}
\affiliation{Department of Mathematics, Trinity College, Hartford, CT 06106, USA}
\email{persebastian.skardal@trincoll.edu}
\author{R. Sevilla-Escoboza}
\affiliation{Centro Universitario de los Lagos, Universidad de Guadalajara, Enrique D\'{i}az de Leon, Paseos de la Monta\~na, Lagos de Moreno, Jalisco 47460, Mexico}
\affiliation{Laboratory of Biological Networks, Center for Biomedical Technology, UPM, Pozuelo de Alarc\'{o}n, 28223 Madrid, Spain}
\author{V. P. Vera-\'Avila}
\affiliation{Centro Universitario de los Lagos, Universidad de Guadalajara, Enrique D\'{i}az de Leon, Paseos de la Monta\~na, Lagos de Moreno, Jalisco 47460, Mexico}
\affiliation{Laboratory of Biological Networks, Center for Biomedical Technology, UPM, Pozuelo de Alarc\'{o}n, 28223 Madrid, Spain}
\author{J. M. Buld\'u}
\affiliation{Laboratory of Biological Networks, Center for Biomedical Technology, UPM, Pozuelo de Alarc\'{o}n, 28223 Madrid, Spain}
\affiliation{Complex Systems Group \& GISC, Universidad Rey Juan Carlos, 28933 M\'ostoles, Madrid, Spain}

\date{\today}

\begin{abstract}
We investigate the existence of an optimal interplay between the natural frequencies of  a group chaotic oscillators and the topological properties of the network they are embedded in. We identify the conditions for achieving phase synchronization in the most effective way, i.e., with the lowest possible coupling strength. Specifically, we show by means of numerical and experimental results that it is possible to define a synchrony alignment function $J(\bm{\omega},L)$ linking the natural frequencies $\bm{\omega_i}$ of a set of non-identical phase-coherent chaotic oscillators with the topology of the Laplacian matrix $L$, the latter accounting for the specific organization of the network of interactions between oscillators. We use the classical R\"ossler system to show that the synchrony alignment function obtained for phase oscillators can be extended to phase-coherent chaotic systems. Finally, we carry out a series of experiments with nonlinear electronic circuits to show the robustness of the theoretical predictions despite the intrinsic noise and parameter mismatch of the electronic components. 
\end{abstract}

\maketitle


{\bf 
The emergence of synchronization of dynamical systems relies on three different aspects that are unavoidably connected:  (i) the dynamical system under study, (ii) the kind of the coupling between dynamical units and (iii) the structure of the network of connections. In this paper we investigate how the distributions of the natural frequencies of a group of dynamical systems that are connected through a certain network topology is crucial to promote or hinder the emergence of phase-synchronization. We show that, given a specific network structure and a set of chaotic oscillators with a certain frequency distribution, there is an optimal allocation for each dynamical system according to its natural frequency. Thus, phase synchronization arises, or not, depending on the interplay between the dynamical and topological properties of the nodes. To verify the robustness of our theoretical predictions, we construct a network of nonlinear electronic circuits and check whether the predicted optimal allocation facilitates the arousal of phase-synchronization. Despite the fact that our results do not apply to non-phase-coherent chaotic oscillators, they demonstrate the existence of an interplay between the dynamical and topological properties of non-identical chaotic systems when trying to achieve strong synchronization.
}

\section{INTRODUCTION}

Synchronization of nonlinear dynamical systems have intrigued scientist in various disciplines studying the emergence of collective phenomena~\cite{pikovsky2001,boccaletti2002}. A large body of research has shown that the particular structure of connections between a system of nonlinear oscillators is crucial in determining the synchronization of the ensemble~\cite{boccaletti2006,arenas2008,newman2010}. Nevertheless, determining effective network structures for a given dynamical system is far from a trivial task. The many classes of dynamical systems, as well as the various types of synchronization we aim to achieve, result in the absence of a unique (or general) network structure maximizing the synchronizability of the system. To this end, it is necessary to understand, for each particular case, how the dynamical system and the structure of connections are intermingled. For example, in the case of heterogeneous phase oscillators such as the Kuramoto model~\cite{acebron2005}, the degree heterogeneity of the network~\cite{restrepo2005} together with the spectral properties of the adjacency and Laplacian matrices~\cite{arenas2008}, can help us to identify what networks are more prone to synchronize and even to asses the time required to reach the synchronization manifold~\cite{almendral2007}.

When the state of the oscillators consist of more than just a single phase, possibly giving rise to chaotic dynamics, synchronization is more complicated. Nevertheless, for an ensemble of identical systems coupled with diffusive coupling, it is possible to obtain a Master Stability Function (MSF) indicating the ability of a system to synchronize by evaluating the stability of the synchronization manifold~\cite{pecora1998}. Extensions of the MSF approach has been proposed for systems with slight parameter mismatch~\cite{sun2009}, but still require that the oscillators are {\it nearly} identical. More recently, it has been pointed out that not only the stability of the synchronized manifold, but the basin of attraction is crucial for the synchronization of the whole system, particularly in real-world scenarios and applications~\cite{menck2013}. The identification of a basin of synchronization is especially useful in real systems where the intrinsic parameter mismatch turns the MSF not applicable. Other approaches have been been developed more recently, allows a network to re-organize itself in the most adequate way. In this scenario it has been demonstrated that the heterogeneity of the nodes can be a driving force behind the evolution of the network structure \cite{kelly2011,scafuti2015}.

Within this framework, a recent work has identified a new perspective in the analysis of synchronization of heterogeneous systems. In Ref.~\cite{skardal2014}, the authors show that given (i) a group of phase oscillators with heterogeneous natural frequencies and (ii) a given network structure, the precise location of the oscillators on the network has crucial consequences on the synchronization of the whole system. In particular, a {\it synchrony alignment function} (SAF) can be defined that describes the interplay between the natural frequencies and the networks structure, and serves as an objective measure of the synchronization of the network. Using the SAF the synchronization properties of a network can be optimized by either (i) strategically allocating oscillators on a fixed network or (ii) tailoring a network to a fixed set of oscillators. This approach accounts for both the heterogeneity in the dynamics of the specific set of oscillators as well as the interactions dictated by the specific network structure. In other words, there is no unique optimal network structure universally valid for all possible sets of frequencies and networks -- in general they have different optimal configurations. Recent work has extended this approach to the case of directed networks~\cite{skardal2016a}, ranking network edges for synchronization~\cite{skardal2016b} and evaluating erosion of synchronization in networks~\cite{skardal2015,skardal2016c}.

In this paper, we translate the concepts introduced in \cite{skardal2014} to the synchronization of chaotic oscillators. We hypothesize that phase-coherent chaotic oscillators with a clear dominant frequency can be treated as phase oscillators and, in turn, the SAF describing the optimal interplay between the natural frequencies and network topology could be used to optimize phase synchronization. Note that this heuristic argument relies on the fact that each dynamical unit is identified with a unique natural frequency, which is not the general case of chaotic oscillators. Nevertheless, this simplification is reasonable for phase coherent systems \cite{farmer1980,rosenblum1996,pikovsky2001} as long as their power spectrum of the dynamical system is narrow enough to clearly identify a dominant frequency. We demonstrate by means of numerical simulations and experiments with nonlinear electronic circuits that the SAF introduced in \cite{skardal2014} is applicable to phase-coherent chaotic oscillators and can be used to find optimal configurations in networks of heterogeneous chaotic oscillators. We also show how the use of the SAF is robust both for random and scale-free topologies but can fail when non-phase-coherent chaotic systems are considered.

The remainder of this paper is organized as follows: In Sec.~\ref{sec2} we describe the oscillator model we are considering and the optimization method using the SAF. In Sec.~\ref{sec3} we present numerical results demonstrating the effectiveness of the optimization method for attaining phase synchronization in networks of chaotic oscillators. In Sec.~\ref{sec4} we validate these numerics by presenting experimental results using networks of nonlinear electronic circuits. In Sec.~\ref{sec5} we conclude with a discussion of our results.

\section{Models and Methods}\label{sec2}

We begin by describing the dynamical system under consideration. Specifically, we study networks of coupled R\"{o}ssler oscillators~\cite{rossler1976}, modified as in Ref.~\cite{leyva2012}. We choose this class of oscillator due to the fact that they represent a paradigmatic example of a class of phase-coherent chaotic oscillators~\cite{pikovsky2001}. Next, we construct a network $\mathcal{G}$ of $N$ R\"{o}ssler oscillators, which are diffusively coupled following the equations:
\begin{align}
\dot{x}_i &= -\alpha_i\left[\Gamma\left(x_i-d\sum_{j=1}^Na_{ij}(x_j-x_i)\right)+\beta y_i + \lambda z_i\right] \label{eqrossler1}\\ 
\dot{y}_i &= -\alpha_i(-x_i+\nu y_i)\\
\dot{z}_i &= -\alpha_i[-g(x_i)+z_i],
\label{eqrossler3}
\end{align}
where
\begin{align}
g(x) = \left\{\begin{array}{rl} 0 & \text{ if }x \le3\\ \mu(x-3),&\text{ if }x>3,\end{array}\right.
\end{align}
is the nonlinear function allowing the oscillators to have a chaotic output.
The parameter $d$ is the global coupling strength, each $\alpha_i$ determines the frequency of oscillator $i$, and the entries $a_{ij}$ define the adjacency matrix $\bf{A}\{a_{ij}\}$ of network $\mathcal{G}$, such that $a_{ij}=1$ if a link exists between oscillators $i$ and $j$, and $a_{ij}=0$ otherwise. Other parameters are $\Gamma = 0.05$, $\beta = 0.5$, $\lambda = 1$, $\mu = 15$, and $\nu = 0.2-10/R$, where $R = 100$. Under this parameter configuration, R\"ossler oscillators have a phase-coherent chaotic output in isolation (i.e., for $d=0$), and their frequencies, $\omega_i$, taken to be the position of the localized peak of the Fourier spectrum, are proportional to $\alpha_i$. (This follows from the fact that, in isolation, the dynamics of oscillator $i$ acts on a timescale of $\alpha_i^{-1}$). The fact that oscillators are phase coherent allow us to evaluate their phase synchronization disregarding the behavior of their amplitude. In particular, we define the phase of oscillator $i$ as the angle represented by the oscillator after projecting onto the $xy$-plane: $\theta_i=\arctan(y_i/x_i)$, as proposed in~\cite{pikovsky1996}. Thus, the degree of phase synchronization of the network can be measured by the order parameter $r=|\sum_{j=1}^Ne^{i\theta_j}|/N$. Here we investigate how the particular frequency of each oscillator is related to the position it occupies in the network and how this relationship affects phase synchronization.

To explore the existence of an optimal frequency-structure interplay, we will adapt the methodological framework introduced in Refs.~\cite{skardal2014,skardal2016a} to the case of phase-coherent chaotic oscillators. In these papers, the authors showed that the degree of phase synchronization in a network of heterogeneous phase oscillators is given by $r\approx1-J(\bm{\omega},L)/2K^2$, where
\begin{align}
J(\bm{\omega},L)=\frac{1}{N}\sum_{j=2}^N\sigma_j^{-2}\langle\bm{u}^j,\bm{\omega}\rangle^2
\label{eqalign}
\end{align}
is the {\it synchrony alignment function} (SAF). Here $\bm{\omega}$ is the vector of the natural frequencies of each oscillator and $L$ is the network Laplacian whose entries are given by $L_{ij}=\delta_{ij}k_i-A_{ij}$. In the case of symmetric adjacency and Laplacian matrices $A$ and $L$, $\sigma_j$ and $\bm{u}^j$ are, respectively, the $j^{\text{th}}$ eigenvalue and associated eigenvector of $L$ \cite{skardal2014}, whereas in the case of asymmetric adjacency and Laplacian matrices $A$ and $L$, $\sigma_j$ and $\bm{u}^j$ are, respectively, the $j^{\text{th}}$ singular value and associated left singular vector of $L$ \cite{skardal2016a}. We include a brief derivation of the SAF in Appendix~\ref{AppA}. Given the setup of the system in Eqs.~(\ref{eqrossler1})--(\ref{eqrossler3}), the heterogeneity of the parameters $\alpha_i$ makes both $A$ and $L$ asymmetric, and therefore we use the asymmetric theory derived in Ref.~\cite{skardal2016a}. 
Given the relationship between $r$ and $J(\bm{\omega},L)$, Eq.~\ref{eqalign} can be used to maximize $r$ (i.e., optimizing phase synchronization), in particular, by minimizing $J(\bm{\omega},L)$ via investigating different configurations of a given set of natural frequencies and/or networks. 

\section{Numerical Simulations: Large networks of phase coherent chaotic oscillators}\label{sec3}

We begin by constructing networks of $N=500$ nodes and $L=1000$ links with and Erd\"os-Renyi random configuration \cite{erdos1959}. Next, we introduce heterogeneity by assigning the pseudo-frequency parameters $\alpha_i$ randomly. Specifically, we obtain each $\alpha_i$ independently from a normal distribution with mean $10$ and standard deviation $0.2$. By doing so, complete synchronization of the whole network can not be achieved, since systems are no longer identical. Instead, we aim to maximize the degree of phase synchronization using the SAF $J(\bm{\omega},L)$ whose input is the distribution of natural frequencies and the Laplacian matrices of the networks.

\begin{figure}[t!]
\centering
\epsfig{file =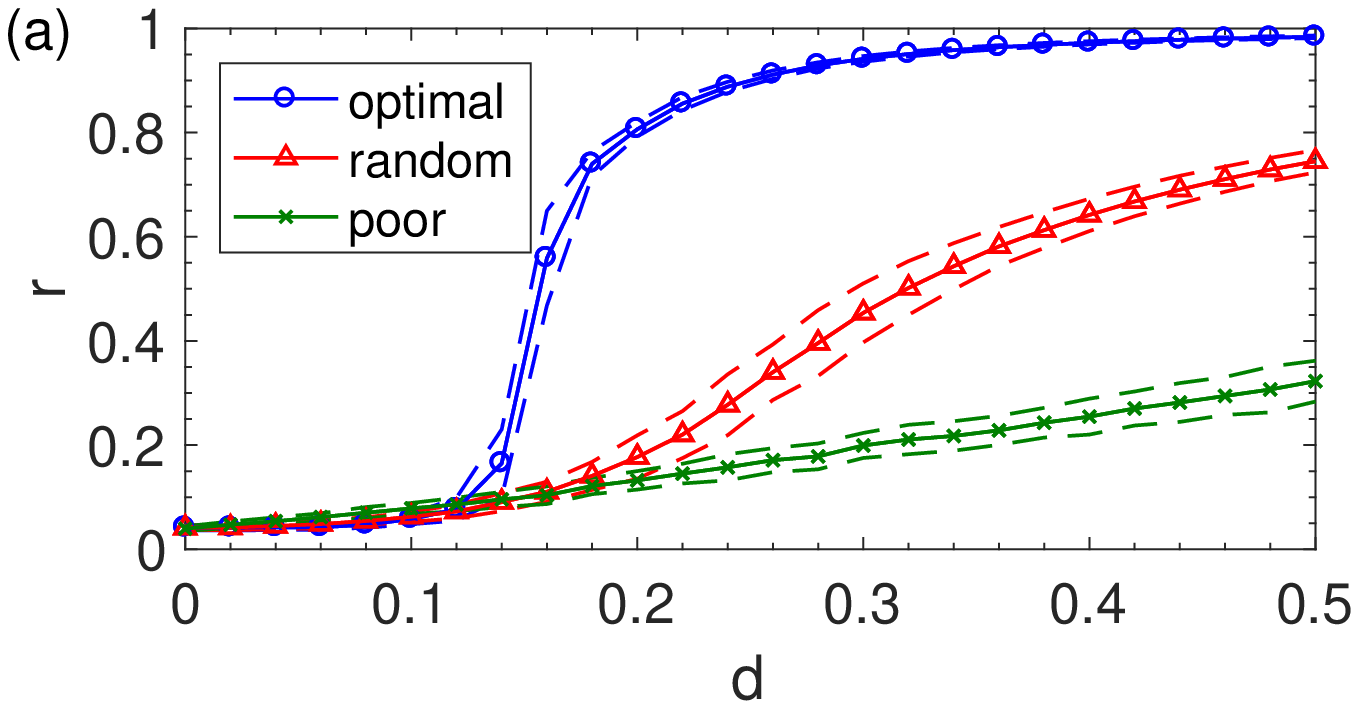, clip =,width=\linewidth }
\epsfig{file =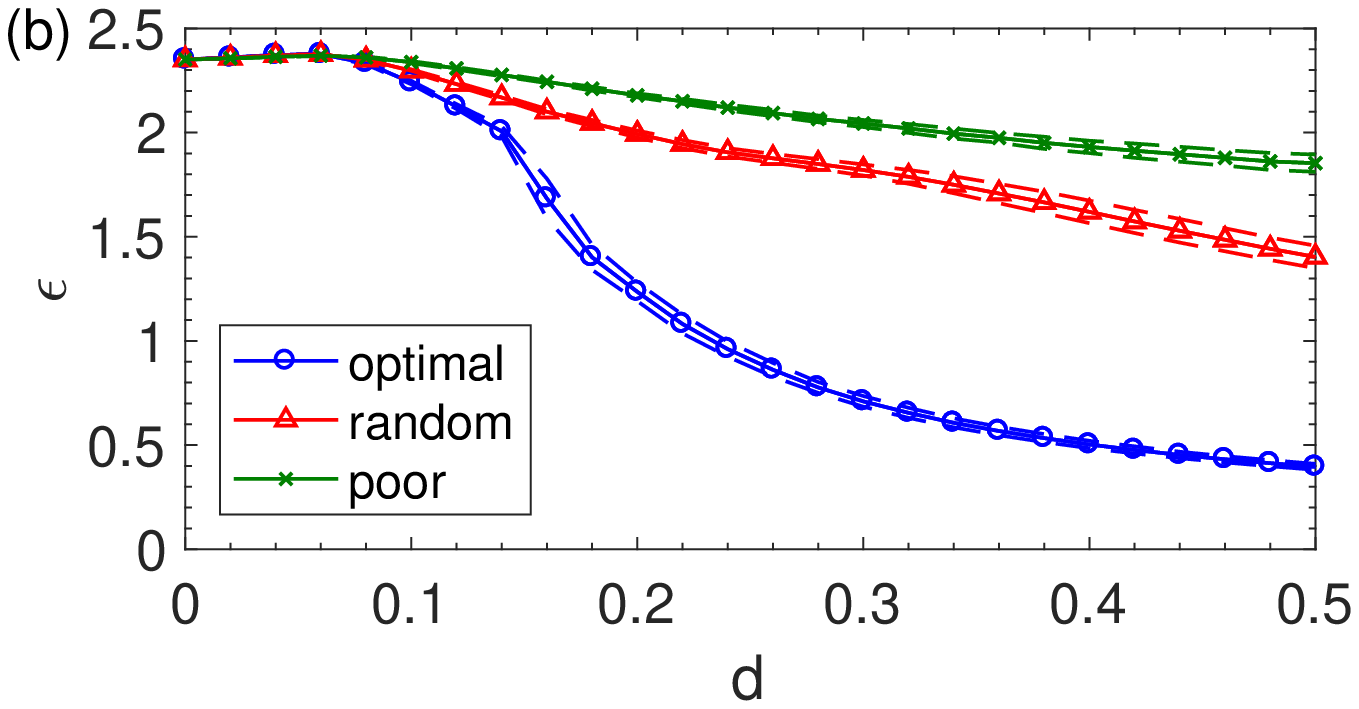, clip =,width=\linewidth }
\caption{(Color online) {\bf Oscillator allocation}. In (a), the order parameter $r$ vs $d$ (coupling strength) for optimal (blue circles), random (red triangles) and poor (green crosses) allocations of oscillators. Solid lines correspond to the average of the simulation of $l=50$ Erd\"os-Renyi random networks of $N=500$ nodes with $L=1000$ links, while the error bars are their corresponding standard deviation. In (b), the corresponding synchronization error.} \label{fig:Rearrange}
\end{figure}

The alignment function is used in two different ways. We first consider {\it oscillator allocation}, where given a fixed network, we seek to arrange frequencies optimally on the network. To do so, we begin by allocating oscillators randomly and implement the following accept-reject algorithm: we propose a switch of two randomly-chosen oscillators, and accept the switch only if the new arrangement yields a smaller value of $J(\bm{\omega},L)$, repeating this process for total number of proposed switches. In Fig.~\ref{fig:Rearrange}(a-b) we plot the results from simulations over a wide range of coupling strengths for networks with optimally allocated frequencies (blue circles), randomly allocated frequencies (red triangles), and poorly allocated frequencies (green crosses). Random allocations are given by the initial network and optimal allocations are obtained after $10^6$ proposed switches. Poor allocations are similarly obtain, but aim to maximize $J(\bm{\omega})$ instead of minimizing it. Solid lines corresponds for the average of $l=50$ different random networks and error bars indicate the standard deviation from the average. We note that optimally allocated set of oscillators performs remarkably well, as shown by the order parameter $r$ (Fig. \ref{fig:Rearrange}a), improving greatly on the random allocation with a sharp transition to a strongly synchronized state near $d\approx0.2$. In addition, the synchronization error, obtained as $\epsilon=\frac{2}{N(N-1)}\sum_{i<j}|x_i-x_j|$ also captures the benefits of the optimal alignment (see Fig. \ref{fig:Rearrange}b), despite the fact that it never vanishes (i.e., achieves perfect synchronization) due to the frequency mismatch of the oscillators. Moreover, phase synchronization can also be mitigated by maximizing $J(\bm{\omega},L)$ as demonstrated by the poorly allocated frequencies (green lines of Fig. \ref{fig:Rearrange}).

\begin{figure}[t!]
\centering
\epsfig{file =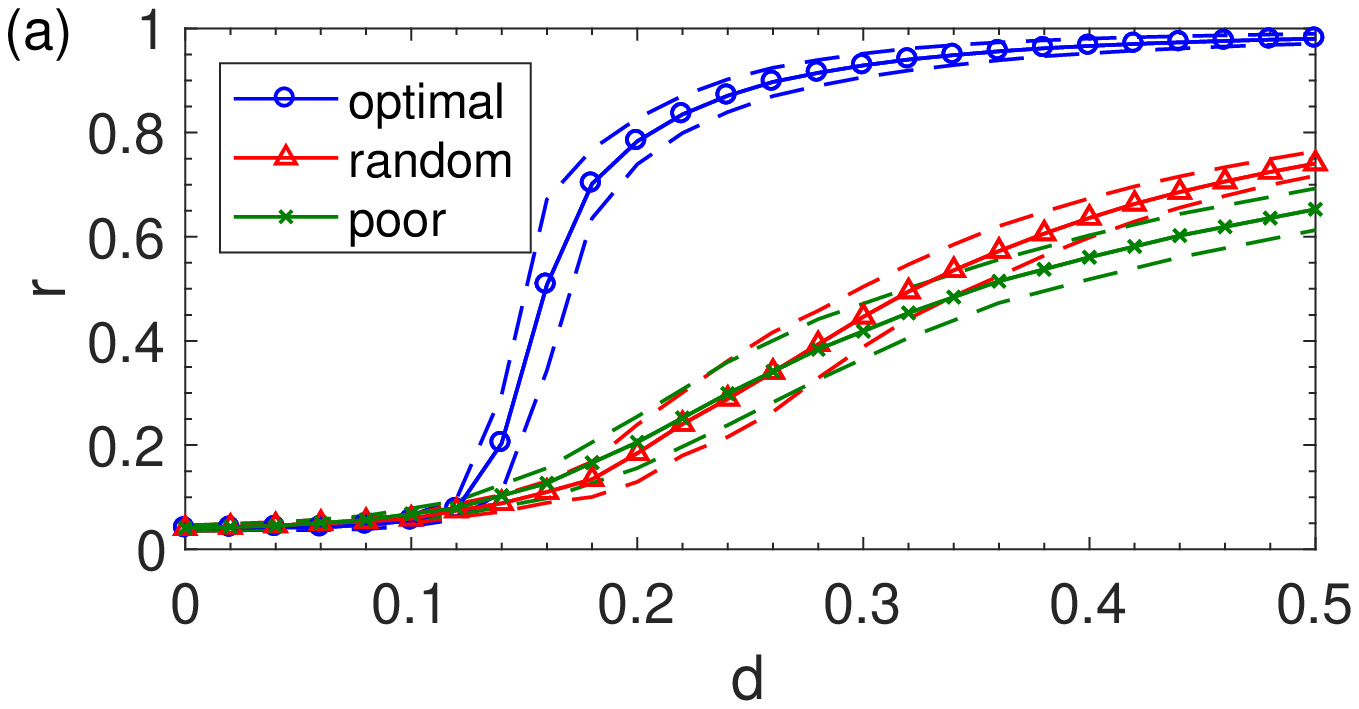, clip =,width=\linewidth }
\epsfig{file =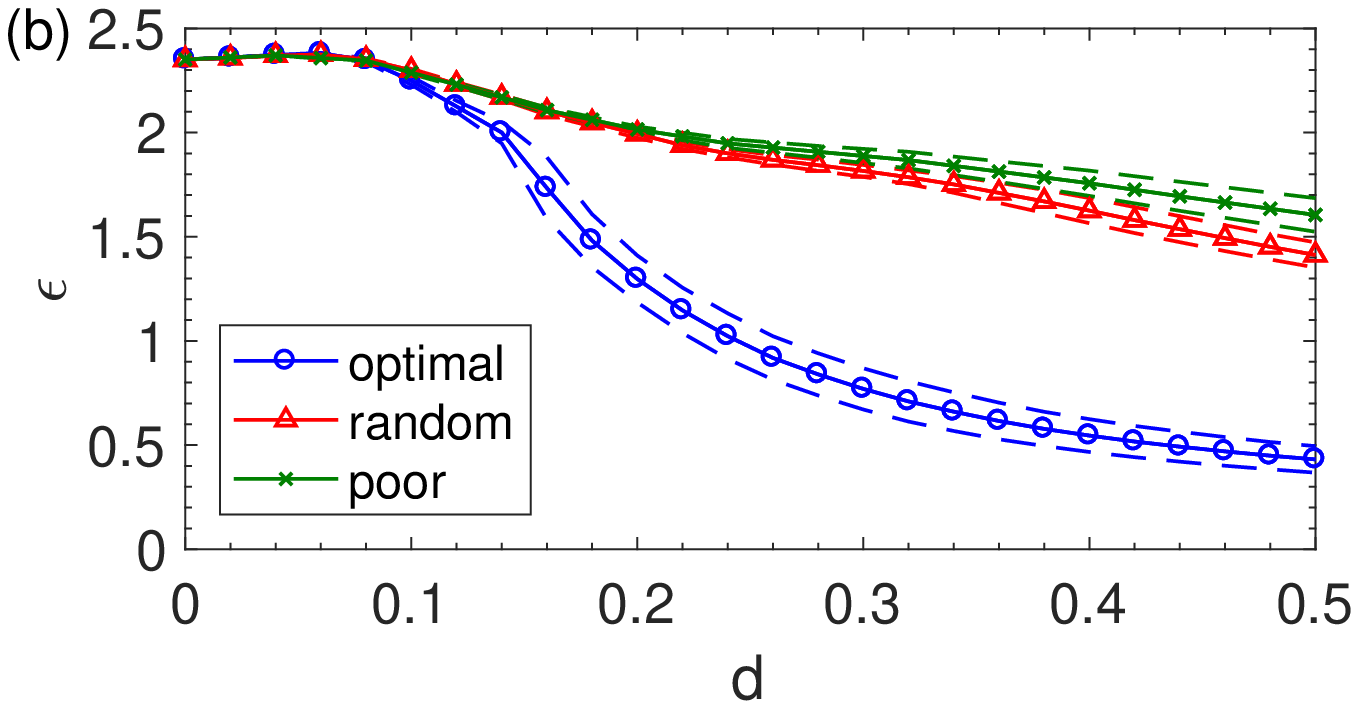, clip =,width=\linewidth }
\caption{(Color online) {\bf Network design}. In (a), the order parameter $r$ vs $d$ (coupling strength) for optimal (blue circles), random (red triangles) and poor (green crosses) network structures. Solid lines correspond to the average of the simulation of $l=50$ Erd\"os-Renyi random networks of $N=500$ nodes with $L=1000$ links, while the error bars are their corresponding standard deviation. In (b), the corresponding synchronization error.}
\label{fig:Rewire}
\end{figure}

Second, we consider a {\it network design} approach, where given a set of oscillators with a predefined frequency distribution, we seek to build an optimal network structure promoting the phase synchronization of the whole ensemble. To do so, we initially allocate oscillators randomly on a network with a random configuration and implement a similar accept-reject algorithm: rather than switching two randomly chosen frequencies, we rewire a randomly chosen link. If the new network yields a smaller $J(\bm{\omega},L)$, we accept the rewiring, and otherwise we reject it. In Fig.~\ref{fig:Rewire}a we plot the results of the order parameter $r$ from simulations over a wide range of coupling strengths for optimal (blue circles), random (red triangles), and poor (green crosses) network structures. Random structures are given by the initial network and optimal structures are obtained after $2\times10^4$ proposed rewirings. Poor structures are similarly obtain, but aim to maximize $J(\bm{\omega})$. As in the previous case, we note that the optimally rewired network performs much better than the random network, displaying a sharp transition to a strongly synchronized state similar to the case of oscillator allocation. The corresponding synchronization error $\epsilon$ also shows that complete synchronization never reached (Fig.~\ref{fig:Rewire}b) due, again, to the heterogeneity of the frequency distribution. Again, phase synchronization can be mitigated by finding poor network structures.

\begin{figure}[t!]
\centering
\epsfig{file =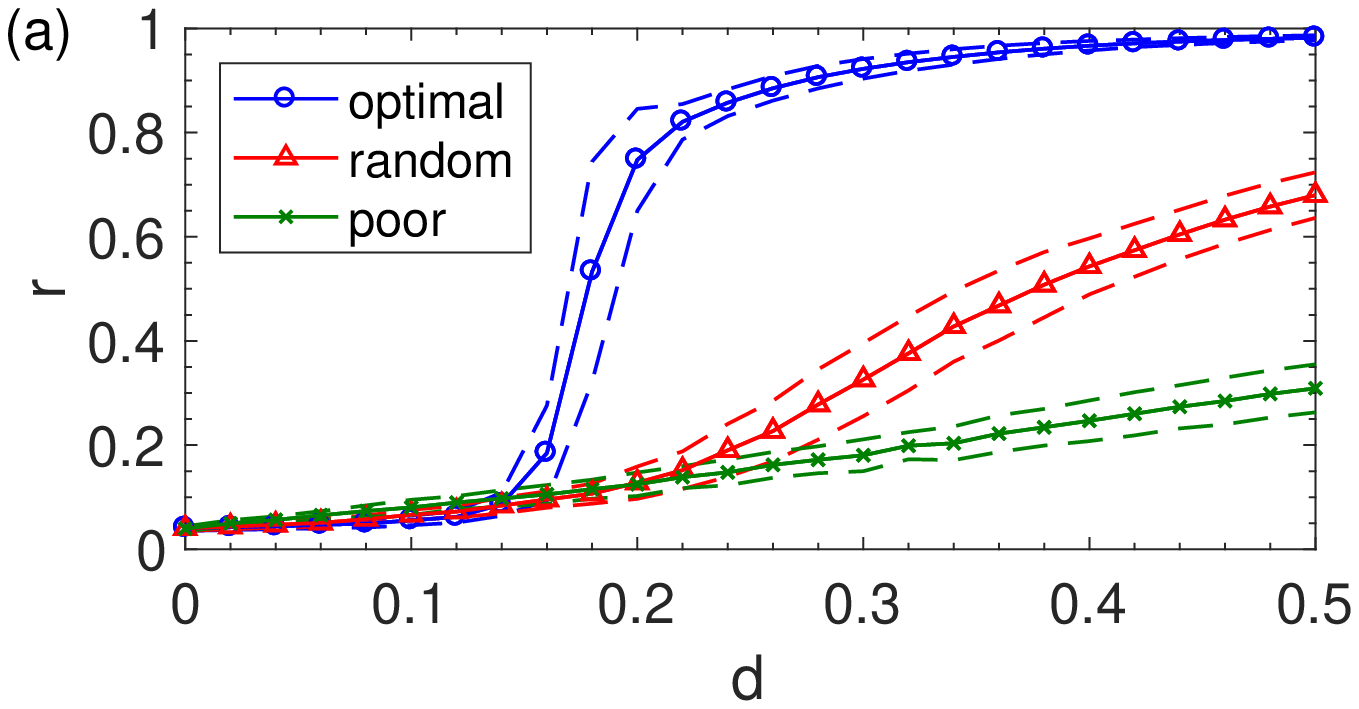, clip =,width=\linewidth }
\epsfig{file =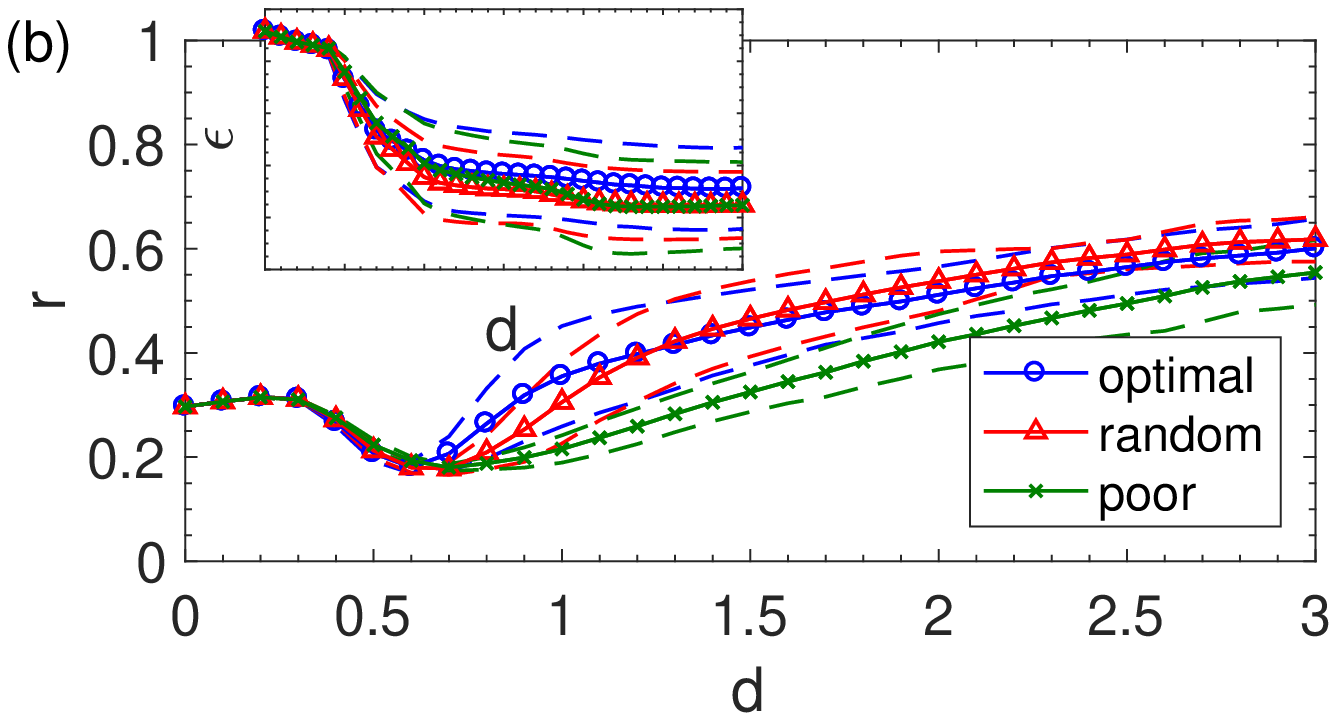, clip =,width=\linewidth }
\caption{(Color online) {\bf Performance of the alignment function in other cases}. In (a), the order parameter $r$ vs $d$ (coupling strength) for optimal (blue circles), random (red triangles) and poor (green crosses) allocations of oscillators on scale-free networks. Solid lines correspond to the average of the simulation of $l=50$ networks with scale-free degree distribution $P(k)\propto k^{-\gamma}$ for $\gamma=3$ and minimum degree $k_0=2$. Error bars indicate standard deviation. In (b), the order parameter $r$ vs $d$ (coupling strength) for optimal (blue circles), random (red triangles) and poor (green crosses) allocations of Lorenz oscillators. Inset: the corresponding synchronization error. Solid lines correspond to the average of the simulation of $l=50$ Erd\"os-Renyi random networks of $N=500$ nodes with $L=1000$ links, while the error bars are their corresponding standard deviation.}
\label{fig:othercases}
\end{figure}

Finally, we explore how the SAF methodology performs in two other cases. In Fig.~\ref{fig:othercases}a we show the order parameter of oscillator allocation when the
underlying network has scale-free structure. Specifically, we consider networks with $N=500$ nodes with scale-free degree distributions $P(k)\propto k^{-\gamma}$ for $\gamma=3$ and minimum degree $k_0=2$, obtained with the configuration model~\cite{bekessy1972}. Similar qualitative results are obtained both for the oscillator allocation and the network reconfiguration (the latter not shown here). In Fig. \ref{fig:othercases}b, the dynamical system has been replaced by a Lorenz system~\cite{lorenz1963,huang2009}, which is a paradigmatic example of a chaotic system that it is not phase-coherent. We summarize the dynamics and phase description of the Lorenz system in Appendix~\ref{AppB}. In this case, the ambiguity in the phase definition leads to a difficulty in defining an characteristic frequency for each oscillator, which in turn hinders the SAF methodology in obtaining optimal configurations of frequencies and network topologies. In particular, we see that networks with ``optimally'' allocated frequencies (blue circles) perform no better than randomly allocated (red triangles) or ``poorly'' allocated (green crosses) frequencies in terms of phase synchronization or the synchronization error (inset).

\section{Experimental results: Implementation with nonlinear electronic circuits}\label{sec4}

\begin{figure}[!bht]
\includegraphics[scale=0.12]{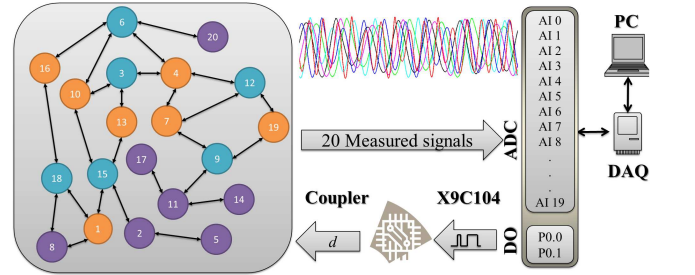}
\caption{ \label{expsetup} {\bf Experimental setup.} On the left, schematic representation of the coupling topology of the 20-circuit network, which can be modified according to the predictions given by the alignment function. The total number of links is always maintained to $L=25$. The coupling strength $d$ is adjusted by means of one digital potentiometers X9C104 controlled by a signal coming from the digital ports P0.0-P0.1 of a DAQ Card. The outputs of the circuit are sent to a set of voltage followers that act as buffers and, then, sent to the analog ports (AI 0 ; AI 1; ... ; AI 19) of the same DAQ Card. The whole experiment is controlled from a PC with Labview 8.5.}
\end{figure}

Next, we validate the robustness of the previous results with a real experiment based on an electronic implementation of a network of R\"ossler oscillators. The experimental design of the whole network, and its control, is shown in Figure \ref{expsetup}. It consists of an electronic array (EA), a multifunction data card (DAQ), and a personal computer (PC). The EA comprises 20 R\"ossler-like electronic circuits (see \cite{sevilla2016} for a detailed description of the electronic schemes) forming a network, whose structure can be modified maintaining the degree of each node. Thus, as a contrast to the numerical results above, the EA is small in size. We translate variables $x_i$, $y_i$ and $z_i$ and all the parameters appearing in Eqs. \ref{eqrossler1}-\ref{eqrossler3} to the three voltages $v_{1i}$, $v_{2i}$ and $v_{3i}$ and a combination of different electronic components, leading to the following circuit equations:
\begin{widetext}
\begin{eqnarray}
v_{1i}(t)&=&-\frac{1}{R_{1}C_{1}}\left(v_{1i} + \frac{R_{1}}{R_{2}}v_{2i}+\frac{R_{1}}{R_{4}}v_{3i} -d\frac{R_{1}}{R_{15}} \sum_{j=1}^{N}{A_{ij} \left[ v_{1j}-v_{1i} \right] } \right)\\
v_{2i}(t)&=&-\frac{1}{R_{6}C_{2}}\left(-\frac{R_{6}R_{8}}{R_{9}R_{7}}v_{1i}+ \left[1- \frac{R_{6}R_{8}}{R_{5}R_{7}}\right]v_{2i}  \right)\\
v_{3i}(t)&=&-\frac{1}{R_{10}C_{3}}\left(-\frac{R_{10}}{R_{11}}G_{v_{1i}}+v_{3i} \right)
\label{eqrosslerExp}
\end{eqnarray}
where
\begin{equation}
G_{x_{i}}= \left\{ \begin{array}{lcc}
0 	&   \text{if}  & x \le Id+Id \frac{R_{14}}{R_{13}}+Vee\frac{R_{14}}{R_{13}} \\
\\ \frac{R_{12}}{R_{14}}x_{i}-Vee\frac{R_{12}}{R_{13}}-Id\left(\frac{R_{12}}{R_{13}}+\frac{R_{12}}{R_{14}} \right)  &   \text{if}  & x> Id+Id \frac{R_{14}}{R_{13}}+Vee\frac{R_{14}}{R_{13}} \\
\end{array}\right.
\label{eqrosslergx}
\end{equation}
\end{widetext}

\begin{figure}[!bht]
\begin{center}
\includegraphics[scale=0.55]{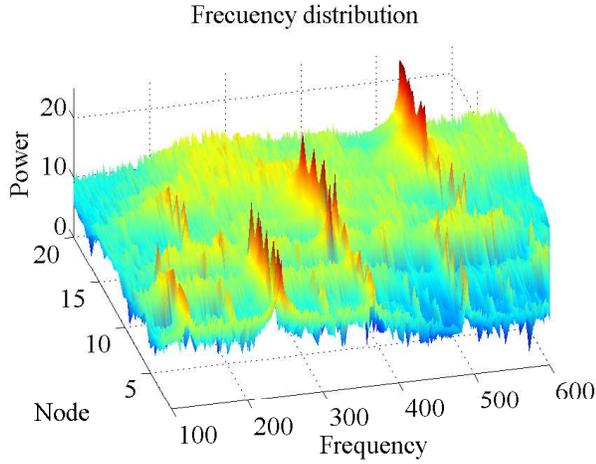}
\end{center}
\caption{\label{Specpatron} {\sf \bf Frequency distribution of the electronic R\"osslers.} Internal capacitances control the frequency of the $N=20$ R\"ossler circuits, taking a value from the set $C_i=\{2.2nF, 3.3nF, 4.7 nF\}$. The ``natural frequency" of each oscillators is obtained from the position of the maximum of their corresponding power spectrum.}
\end{figure}

The values of all electronic components are summarized in Tab. \ref{TCexp}. Importantly, all chaotic oscillators have the same internal parameters with the exception of the capacitances $C_{i}$ that, in turn, define the natural frequency of oscillation of each unit. This way, capacitances take different values for the set $C_{i}=\{2.2nF, 3.3nF, 4.7 nF\}$, leading to a distribution of frequencies within the interval $(240 - 540)$ Hz. The relation between $\alpha_i$
 of the theoretical model and the capacitances is given by  $\alpha _{v_{1i}}=\frac{1}{R_{4}C_{1}}$, 
 $\alpha _{v_{2i}}=\frac{R_{8}}{R_{7}R_{9}C_{2}}$ and $\alpha _{v_{3i}}=\frac{1}{C_{3}R_{10}}$. As a consequence, we obtain an ensemble of oscillators 
 whose natural frequencies $\omega_i$, in isolation, are inversely proportional to the value of the capacitances. Note that, due to the tolerance of the electronic components (between $5\%$ and $10\%$), the frequencies of the oscillators also suffer a dispersion that goes beyond the nominal values of the capacitances. Figure \ref{Specpatron} shows the power spectrum of the $N=20$ oscillators for $d=0$, i.e., in isolation. Each R\"ossler circuit has an individual electronic coupler controlled by a digital potentiometer (XDCP), which is adjusted by a digital signal coming from ports P0.0-1 (see Fig. \ref{expsetup}). The digital port P0.0 is used to set the value of the coupling resistance ($d$), while port P0.1 increases/decreases the resistance of a voltage divisor controlling the final coupling strength $d$ (and allowing to test 100 discretized values of $d$). All the experimental process is controlled by a virtual interface developed in Labview, which can be considered as a state machine. This way, the experimental procedure is as follows: first, $d$ is set to zero, after a waiting time of 500 ms (roughly corresponding to $P=(120 - 270)$ cycles of the autonomous systems), the signals corresponding to the $x(t)$ variables of the 20 circuits are acquired by the analog ports ( AI 0 ; AI 1; ... ; AI 19). Once this is recorded, the value of $\sigma$ is increased by one step, and the process is repeated until the maximum value of $\sigma$ is reached.
\begin{table}[h] \centering
\begin{tabular}{|c|c|c|c|c|c|}
\hline \hline
$C_{1-3}$			& $2.2$nF 				& $3.3$nF 				&$4.7$nF  					\\ 
\hline \hline
$R_1 = 2M\Omega$	& $R_2 = 200K\Omega$ 	&$R_3 = 10K\Omega$ 		& $R_4 = 100K\Omega$ 		 \\ 
\hline
$R_5 = 50K\Omega$	& $R_6 = 5MK\Omega$ 	&$R_7 = 100K\Omega$ 	& $R_8 = 10K\Omega$ 		 \\ 
\hline
$R_9 = 10K\Omega$	& $R_{10}= 100K\Omega$ 	&$R_{11}= 100K\Omega$ 	& $R_{12}= 150K\Omega$ 		  \\
\hline
$R_{13}= 68K\Omega$	& $R_{14}= 10K\Omega$ 	&$R_{15}= 100K\Omega$ 	& $R_C= R_3+R_5$ 		 \\ 
\hline
$Id = 0.7$			& $Vee = 15$ 			&$d = [0-0.6]$ 		&  		 \\ 
\hline \hline
\end{tabular}
\caption{Values of the electronic components used for the construction of the electronic version of the R\"ossler system.}
\label{TCexp}
\end{table}

Once the $x_i(t)$ variable of all circuits is recorded, we obtain the equivalent instantaneous phase as $\phi_i(t)=\ 2 \pi l_i + 2 \pi \frac{t-t_{l_i}}{t_{l_i+1}-t_{l_i}}$, for each interval $t_{l_i} \leq t < t_{l_i+1}$, where $t_{l_i}$ and $t_{l_i+1}$ are the times of to consecutive maxima of the variable $x_i(t)$ \cite{pikovsky2001}.
It is important to remark that, for phase coherent systems, such a measure of instantaneous phase is fully equivalent to the geometrical phases used in the numerical simulations \cite{arenas2008}.

\begin{figure}[!bht]
\begin{center}
\epsfig{file =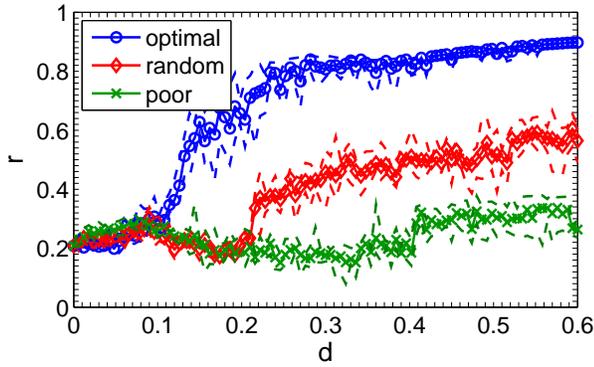, clip =,width=\linewidth }
\end{center}
\caption{\label{drexp} {\sf \bf Experimental test of the optimal alignment.} Order parameter $r$ of a network of heterogeneous R\"ossler circuits as a function
of the coupling strength $d$ and for a given network structure (see Fig.  \ref{expsetup} for a representation of the network). Three different allocations are shown: the optimal alignment predicted by Eq. \ref{eqalign} (blue circles), a random configuration (red diamonds) and a poor allocation (green crosses). Dashed lines are the corresponding standard deviations obtained from five different experimental realizations.}
\end{figure}


\begin{figure}[!bht]
\begin{center}
\epsfig{file =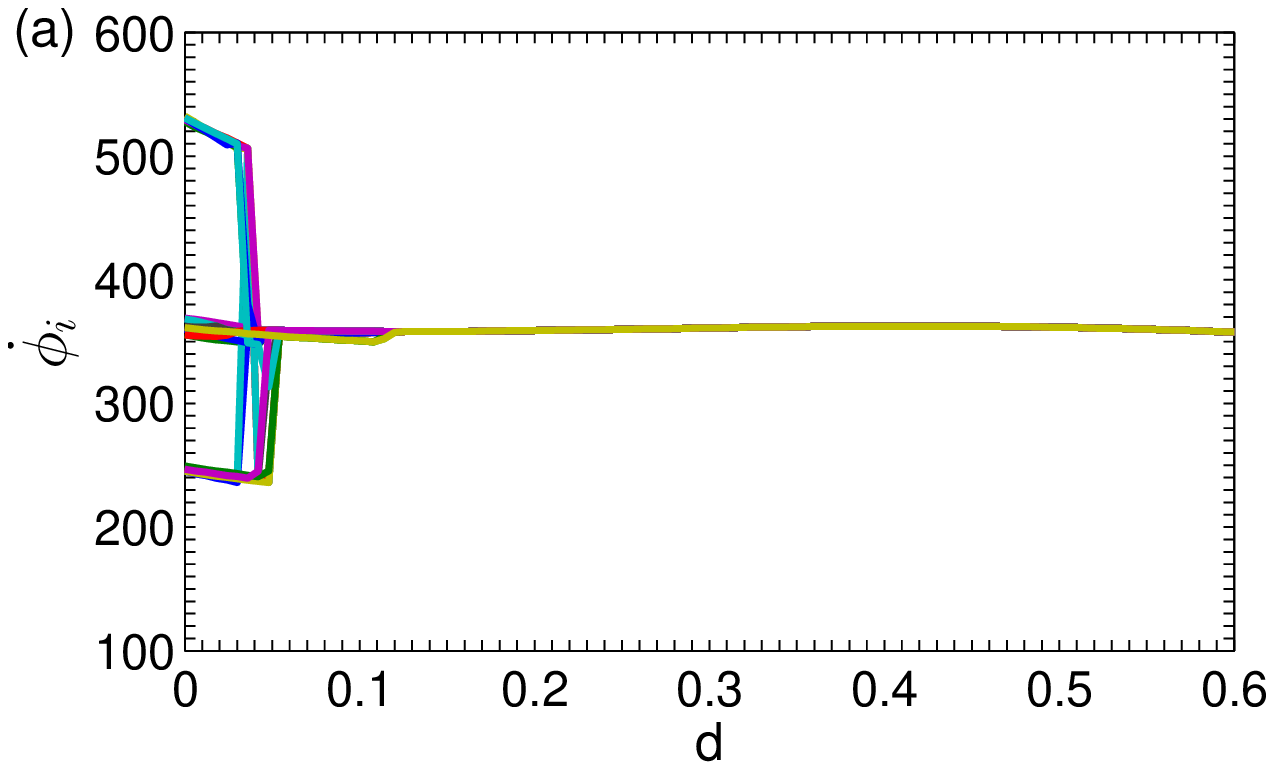, clip =,width=\linewidth }
\epsfig{file =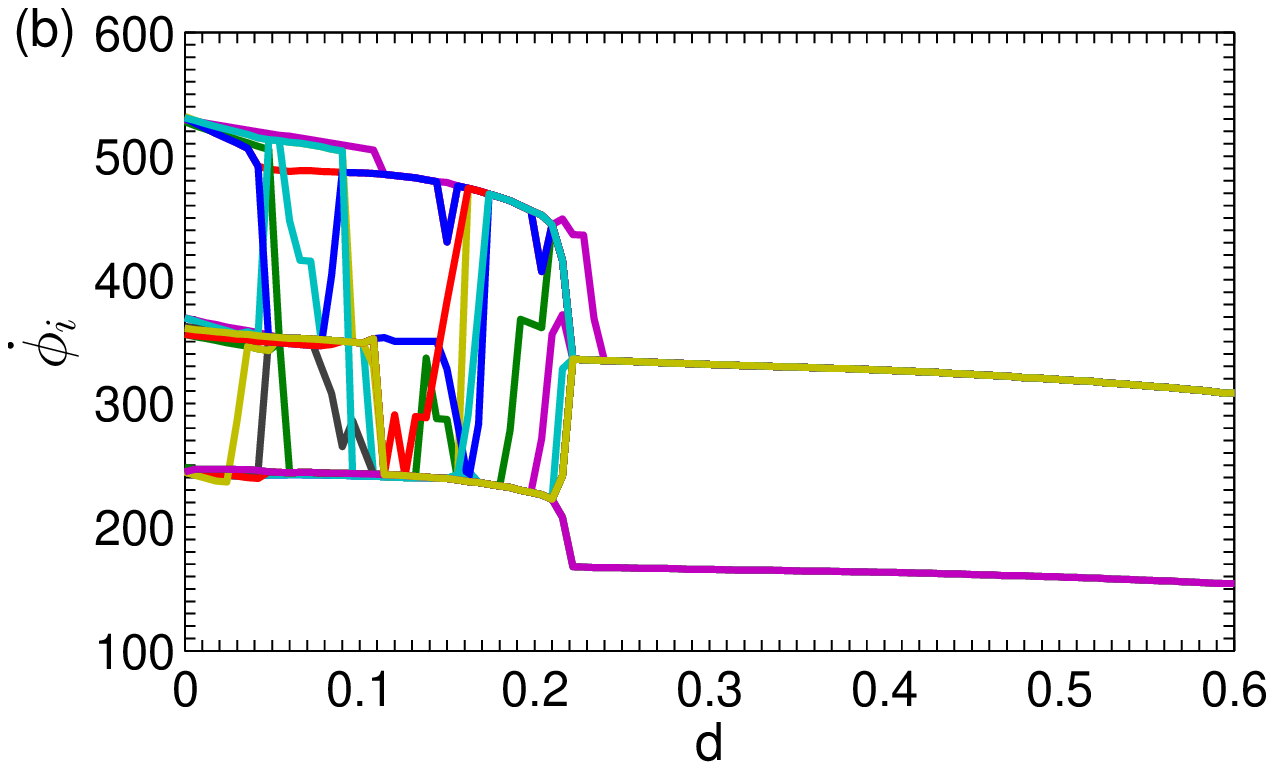, clip =,width=\linewidth }
\epsfig{file =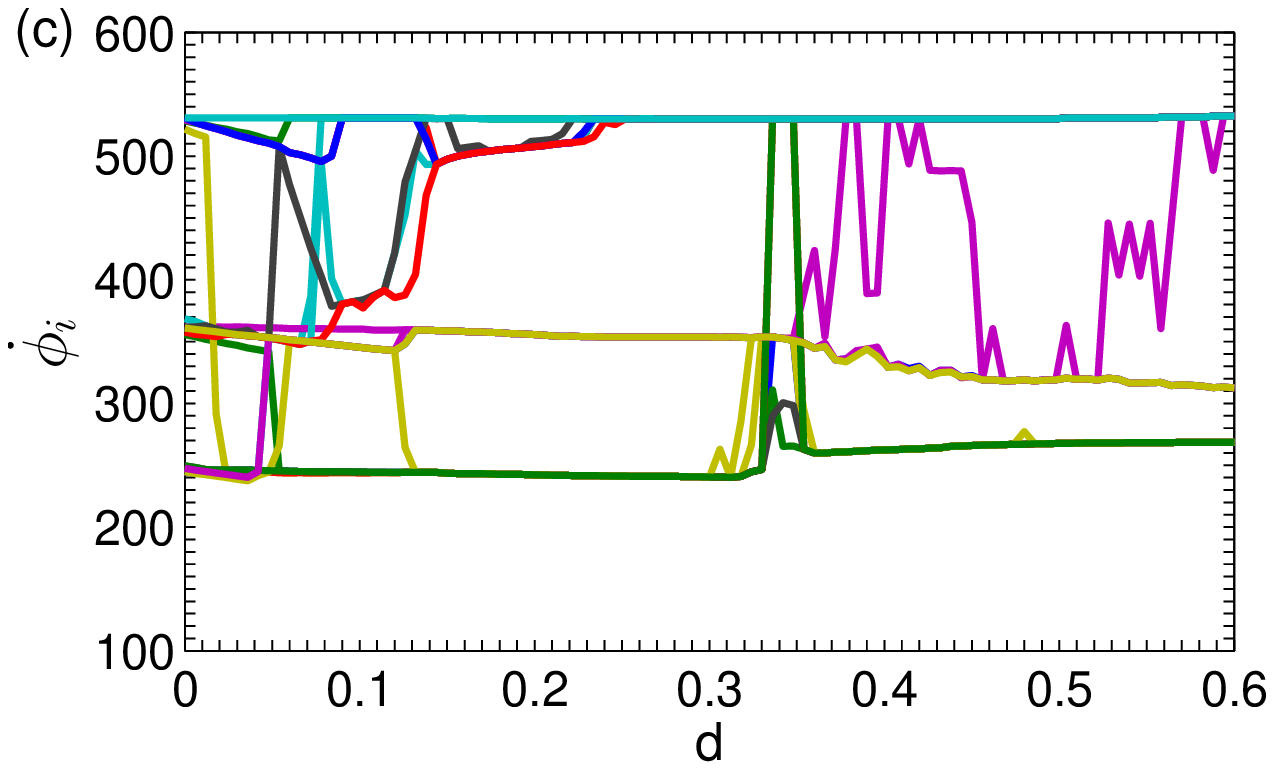, clip =,width=\linewidth }
\end{center}
\caption{\label{FrecTran} {\sf \bf Instantaneous frequency vs coupling strength $d$ for different network alignments.}  In all figures,
we plot the position of the highest peak of the power spectrum of each oscillator as a function of $d$. In (a), the optimal alignment
leads to a frequency locking for couplings strengths higher than $d \sim 0.5$. In b), a random allocation of the oscillators leads to the formation
of synchronization clusters even for high values of the coupling strength. Finally, in (c), the poor allocation results in the coexistence of synchronization clusters together with a frequency drift of certain oscillators.}
\end{figure}

Next, we evaluate the performance of the theoretical predictions given by the alignment function $J(\bm{\omega},L)$. We construct the specific network structure is shown in Fig. \ref{expsetup}, i.e. $N=20$ R\"ossler oscillators with the frequency distribution shown in Fig. \ref{Specpatron} and with a fixed average degree of $\langle k \rangle=2.5$. Next, we obtain the synchronization parameter $r$ vs the coupling strength for three different configurations: (i) the optimal allocation according to an alignment function whose input are the average frequency of the oscillators and the Laplacian matrix of the network; (ii) a random allocation of the oscillators on the network  and finally (iii) a poor allocation (i.e., maximizing the alignment function). Figure \ref{drexp} shows how the optimal allocation allows the ensemble of oscillators to synchronize at lower coupling strengths and a the same time, to reach higher values of the order parameter. A further inspection of the oscillators' instantaneous frequency reveals the path to synchronization for both the optimal and random allocations. Figure \ref{FrecTran} shows the instantaneous frequency of the $N$ oscillators, indicated by the position of the highest peak of their corresponding power spectrum, as the coupling strength $d$ is increased. When the optimal alignment is implemented (Fig. \ref{FrecTran}(a)), we observe a frequency locking occurring at $d \sim 0.5$, which precedes the subsequent phase locking of the whole system reached for values of $d>0.1$ (see blue line in Fig. \ref{drexp}). Interestingly, Figs. \ref{FrecTran}(b)-(c) show that deviations from the optimal alignment lead to, respectively, b) the formation of frequency locked clusters preventing the whole system to synchronize and c) the drift of a series of oscillators along the dominant frequencies of the system. In this way, we can observe that, in the case of random allocation, we obtain two different clusters of synchronization, which hold even for large values of $d$ and prevent to achieve the phase synchronization of the whole system (Fig. \ref{drexp}(b)). In the case of the network with the poor allocation, the situation is even worse, since a series of oscillators drift from cluster to cluster leading to an even lower value of the order parameter. It is worth mentioning that, in \cite{skardal2014}, a positive correlation between $\omega_i$ and $k_i$ was reported. However, such correlations are not observed in our experiments, which can be attributed to the finite size effects related to the experimental limitations resulting in a low heterogeneity in the number of nodes ($N=20$), natural frequencies (only three different frequencies are implemented) and node degrees ($1\leq k \leq 4$).


\section{Conclusions}\label{sec5}


Our results show, both numerically and experimentally, that when aiming to synchronize a network of phase-coherent chaotic systems, there exists an optimal alignment between the frequency of the oscillators and the topological properties of the network. The existence of an optimal correlation is shown from two perspectives: (i) promoting synchronization by the correct allocation of a group of heterogeneous oscillators over a given network structure and (ii) allowing the network to reorganize according to an alignment function. In both cases, the alignment function introduced in \cite{skardal2014} for Kuramoto oscillators correctly defines the optimal frequency-topology correlation for R\"ossler systems with phase-coherent chaotic behavior. 

At the same time, our results raise a series of questions to be addressed in the future. First, we have seen that the application of the alignment function to the Lorenz system does not give successful results. This fact suggests that the alignment function of chaotic oscillators that are not phase-coherent needs to include, somehow, the complexity of the dynamics and not only the dominant frequency of the system. More generally, it raises the question of how to define new alignment functions accounting for power spectrums with large dispersion. Is it possible to define a specific alignment function to each considered dynamical system? What is the relation between the optimal topologies for different dynamical systems? Both questions require significant attention in the near future and might be addressed in part by blend the SAF approach used here with of other approaches~\cite{Gottwald2015,Pinto2015}. In addition, it would be interesting to carry out experiments with a number of nodes two or three orders of magnitude higher, which would  allow to 
 check the theoretical predictions regarding the generality of our results in scale-free networks (which requiere a high number of nodes) and the expected correlations
 between the degree of the nodes and their natural frequencies.

On the other hand, there is a pristine field concerned about the identification of optimal topologies in real systems. For example, in brain networks, recent works have focused on the interplay between the dynamical and topological properties of different brain regions, and how the underlying physical connections constrain the functional networks associated to different cognitive and motor tasks \cite{simas2015,stam2016,garces2016,battiston2016}. In this respect, the identification of alignment functions could help to understand the interplay between structural and functional networks. How close the actual configuration of a functional network is from the optimal topology and how a neurodegenerative disease may alter the possible alignment between dynamical and structural properties would be promising applications.

Authors acknowledge D. Papo, R. Guti\'errez and P.L. del Barrio for fruitful conversations. J.M.B. is founded by MINECO (FIS2013-41057-P). R.S.E. acknowledges Universidad de Guadalajara, Culagos (Mexico) for financial support (PIFI 522943 (2012) and Becas Movilidad 290674-CVU-386032). P.S.S. acknowledges financial support from the James S. McDonnell Foundation.

\begin{appendix}

\section{Derivation of the Synchrony Alignment Function}\label{AppA}

In this Appendix we present a short derivation of the synchrony alignment function (SAF) as first described in Ref.~\cite{skardal2014}. We consider a network of $N$ phase oscillators evolving according to
\begin{align}
\dot{\theta}_i=\omega_i+d\sum_{j=1}^Na_{ij}H(\theta_j-\theta_i),\label{eq:App01}
\end{align}
where $H$ is a coupling function which we assume satisfies $H'(0)>0$ and $|H(0)/\sqrt{2}H'(0)|\ll1$ to ensure the possibility of synchronization (this last condition ensures that the coupling {\it frustration} is sufficiently small~\cite{skardal2015}). We then consider the {\it strong coupling regime} where a strongly synchronized state, i.e., $r\approx1$, can be attained such that the difference between any two network-adjacent phases is small, $|\theta_j-\theta_i|\ll1$. In this scenario Eq.~(\ref{eq:App01}) can be linearize to
\begin{align}
\dot{\theta}_i\approx\tilde{\omega}_i-\tilde{d}\sum_{j=1}^NL_{ij}\theta_j,\label{eq:App02}
\end{align}
where $\tilde{\omega}_i=\omega_i+KH(0)k_i$ and $\tilde{d}=dH'(0)$. Note that in the case of Kuramoto coupling where $H(\theta)=\sin\theta$ we have that $\tilde{\omega}_i=\omega_i$ and $\tilde{d}=d$. Dropping the $\sim$-notation, we next enter the rotating frame $\theta\mapsto\theta+\Omega t$ where $\Omega$ is the collective frequency variation~\cite{skardal2016d}. The stationary solution of Eq.~(\ref{eq:App02}) can then be written in vector form
\begin{align}
\bm{\theta^*}=L^\dagger\bm{\omega}/d,\label{eq:App03}
\end{align}
where $L^\dagger$ is the Moore-Penrose pseudoinverse of the Laplacian $L$~\cite{BenIsrael}. In particular, if $L$ is symmetric its pseudoinverse is defined by its eigen decomposition, $L^\dagger=\sum_{j=2}^N\sigma_j^{-1}\bm{u}^j\bm{u}^{jT}$, where $\sigma_j$ and $\bm{u}^j$ are, respectively, the $j^{\text{th}}$ eigenvalue and associated eigenvector of $L$. However, if $L$ is asymmetric, its pseudoinverse is defined by its singular value decomposition, $L^\dagger=\sum_{j=2}^N\sigma_j^{-1}\bm{u}^j\bm{v}^{jT}$, where $\sigma_j$, $\bm{u}^j$, and $\bm{v}^j$ are, respectively, the $j^{\text{th}}$ singular value, associated left singular vector, and associated right singular vectors of $L$. Finally, noting that to leading order the order parameter can be written
\begin{align}
r\approx1-\frac{\|\bm{\theta^*}\|^2}{2N},\label{eq:App04}
\end{align}
we insert Eq.~(\ref{eq:App03}) into Eq.~(\ref{eq:App04}) to obtain
\begin{align}
r\approx1-J(\bm{\omega},L)/2K^2,\label{eq:App05}
\end{align}
where
\begin{align}
J(\bm{\omega},L)=\frac{1}{N}\sum_{j=2}^N\sigma_j^{-2}\langle\bm{u}^j,\bm{\omega}\rangle^2\label{eq:App06}
\end{align}
is the SAF presented in the main text in Eq.~(\ref{eqalign})

\section{Description of the Lorenz dynamics}\label{AppB}

In this Appendix we briefly describe the dynamics and phase description of the Lorenz oscillators used in the main text. Given a network of $N$ nodes we consider the following system:
\begin{align}
\dot{x}_i&=\alpha_i\left[\sigma(x_i-y_i)+d\sum_{j=1}^Na_{ij}(x_j-x_i)\right]\\
\dot{y}_i&=\alpha_i\left[x_i(\rho-z_i)-y_i\right]\\
\dot{z}_i&=\alpha_i\left[x_iy_i-\beta z_i\right].
\end{align}
As in the case of the R\"{o}ssler oscillators, $d$ represents a global coupling parameter, the entries $a_{ij}$ describe the network structure, and $\alpha_i^{-1}$ determines the timescale of oscillator $i$. Other parameters are chosen $\sigma=10$, $\rho=28$, and $\beta=8/3$ to induce a chaotic state. The topology of the chaotic Lorenz attractor, however, is not phase-coherent making a phase $\theta_i$ difficult to extracted from the state $(x_i,y_i,z_i)^T$. Here we define a phase, first by defining the variable $u_i=\sqrt{x_i^2+y_i^2}$, then defining $\theta_i$ as the angle made by the state $(u_i,z_i)$ around the unstable fixed point $(u^*,z^*)=\left(\sqrt{2\beta(\rho-1)},\rho-1\right)$, i.e.,
\begin{align}
\theta_i=\arctan\left(z_i-(\rho-1),u_i-\sqrt{2\beta(\rho-1)}\right).
\end{align}
Using this formalism of each phase $\theta_i$, the degree of phase synchronization is then given by the same order parameter as defined in the main text.

\end{appendix}


\begin{thebibliography}{}

\bibitem{pikovsky2001}
A. Pikovsky, M. Rosemblum and J. Kurths, {\it ``Synchronization: A universal concept in nonlinear sciences''},
Cambridge University Press, Cambridge, U.K. (2001).

\bibitem{boccaletti2002} S. Boccaletti, J. Kurths, G. Osipov, D.L. Valladares and C.S. Zhou, 
``The synchronization of chaotic systems",
{\it Phys. Rep.} {\bf 366}, 1-101 (2002).

\bibitem{boccaletti2006} S. Boccaletti, V. Latora, Y. Moreno, M. Chavez and D.U. Hwang, 
{\it Phys. Rep.} {\bf 424},175-308 (2006). 

\bibitem{arenas2008} A. Arenas, A. D\'{\i}az-Guilera, J. Kurths, Y. Moreno and C. Zhou, 
``Synchronization in complex networks",
{\it Phys. Rep.} {\bf 469}, 93-153 (2008).

\bibitem{newman2010} M.E.J. Newman,
{\it Networks: An introduction.} Oxford University Press, New York (2010).

\bibitem{acebron2005} J.A. Acebr\'on, L.L. Bonilla, C.J. P\'erez-Vicente, F. Ritort, and R. Spigler,
``The Kuramoto model: A simple paradigm for synchronization phenomena",
{\it Rev. Mod. Phys.} {\bf 77}, 137-185 (2005). 

\bibitem{restrepo2005} J.G. Restrepo, E. Ott and B.R. Hunt, 
``Onset of synchronization in large networks of coupled oscillators", 
{\it Phys. Rev. E } {\bf 71}, 036151 (2005).

\bibitem{almendral2007} J.A. Almendral and A. D\'iaz-Guilera,
``Dynamical and spectral properties of complex networks",
{\it New Journal of Physics} {\bf 9}, 187 (2007).

\bibitem{pecora1998} L.M. Pecora and T.L. Carroll,
``Master Stability Functions for Synchronized Coupled Systems'', 
{\it Phys. Rev. Lett.} {\bf 80}, 2109 (1998).

\bibitem{sun2009} J. Sun, E.M. Bollt and T. Nishikawa,
``Master stability functions for coupled nearly identical dynamical systems",
{\it Europhys. Lett.} {\bf 85}, 60011 (2009).

\bibitem{menck2013} P.J. Menck, J. Heitzig, N. Marwan and J. Kurths,
``How basin stability complements the linear-stability paradigm",
{\it Nat. Phys.} {\bf 9}, 89-92 (2013).

\bibitem{kelly2011} D. Kelly and G. A. Gottwald,
``On the topology of synchrony optimized networks of a Kuramoto-model with non-identical oscillators",
{\it Chaos} {\bf 21}, 025110 (2011).

\bibitem{scafuti2015} F. Scafuti, T. Aoki and M. di Bernardo,
``Heterogeneity induces emergent functional networks for synchronization",
{\it Phys. Rev. E} {\bf 91}, 062913 (2015).

\bibitem{skardal2014} P. S. Skardal, D. Taylor and J. Sun,
``Optimal synchronization of complex networks",
{\it Phys. Rev. Lett.} {\bf 113}, 144101 (2014).

\bibitem{skardal2016a} P. S. Skardal, D. Taylor and J. Sun,
``Optimal synchronization of directed complex networks",
{\it Chaos} {\bf 26}, 094807 (2016).

\bibitem{skardal2016b} D. Taylor, P. S. Skardal and J. Sun,
``Synchronization of heterogeneous oscillators under network modifications: Perturbations and optimization of the synchrony alignment function",
{\it SIAM J. Appl. Math.} {\bf 76}, 1984 (2016).

\bibitem{skardal2015} P. S. Skardal, D. Taylor, J. Sun and A. Arenas,
``Erosion of synchronization in networks of coupled oscillators",
{\it Phys. Rev. E} {\bf 91}, 010802(R) (2015).

\bibitem{skardal2016c} P. S. Skardal, D. Taylor, J. Sun and A. Arenas,
``Erosion of synchronization: Coupling heterogeneity and network structure",
{\it Physica D} {\bf 323--324}, 40 (2016).

\bibitem{farmer1980} D. Farmer, J. Crutchfield, H. Froehling, N. Packard and R. Shaw, 
``Power spectra and mixing properties of strange attractors",
{\it Ann. New York Acad. Sci.} {\bf 357}, 453 (1980).

\bibitem{rosenblum1996} M.G. Rosenblum, A.S. Pikovsky and J. Kurths,
``Phase Synchronization of Chaotic Oscillators",
{\it Phys. Rev. Lett.} {\bf 76}, 1804 (1996).


\bibitem{rossler1976} O.E. R{\"o}ssler, 
``An equation for continuous chaos'',
{\it Phys. Lett. A} {\bf 57}, 397 (1976).

\bibitem{leyva2012} I. Leyva, R. Sevilla-Escoboza, J. M. Buld\'u, I. Sendi\~na-Nadal, J. Gomez-Garde\~nes, 
A. Arenas, Y. Moreno, S. Gomez, R. Jaimes-Re\'ategui and S. Boccaletti, 
``Explosive First-Order Transition to Synchrony in Networked Chaotic Oscillators", 
{\it Phys. Rev. Lett.} {\bf 108}, 168702 (2012).


\bibitem{pikovsky1996} A.S. Pikovsky, M.G. Rosenblum and J. Kurths, 
``Synchronization in a population of globally coupled chaotic oscillators",
{\it Europhys. Lett.} {\bf 34}, 165-170 (1996).
 
\bibitem{erdos1959} P. Erd\"os and A. R\'enyi,
``On Random Graphs. I", 
{\it Publicationes Mathematicae} {\bf 6}, 290-297 (1959).

 \bibitem{bekessy1972}  A. Bekessy, P. Bekessy and J. Komlos,
 ``Asymptotic enumeration of regular matrices",
{\it Stud. Sci. Math. Hung.} {\bf 7}, 343 (1972).
 
 \bibitem{lorenz1963} E.N. Lorenz,
 ``Deterministic nonperiodic flow",
 {\it J. Atmos. Sci.} {\bf 20}, 130-141 (1963).

\bibitem{huang2009} L. Huang, Q. Chen, Y.-C. Lai and L.M. Pecora,
``Generic behavior of master-stability functions in coupled nonlinear dynamical systems",
{\it Phys. Rev. E} {\bf 80}, 036204 (2009).

\bibitem{sevilla2016} R. Sevilla-Escoboza and J.M. Buld\'u,
``Synchronization of networks of chaotic oscillators: Structural and dynamical datasets",
{\it Data in Brief}, {\bf 7}, 1185-1189 (2016). 

\bibitem{Gottwald2015} G. Gottwald,
``Model reduction for networks of coupled oscillators",
{\it Chaos} {\bf 25}, 053111 (2015).

\bibitem{Pinto2015} R. S. Pinto and A. Saa,
``Optimal synchronization of Kuramoto oscillators: A dimensional reduction approach",
{\it Phys. Rev. E} {\bf 92}, 062801 (2015).

\bibitem{simas2015} T. Simas, M. Chavez, P.R. Rodr\'iguez and A. D\'iaz-Guilera,
``An algebraic topological method for multimodal brain networks comparisons",
{\it Front. Psychol.} {\bf 6}, 904 (2015).

\bibitem{stam2016} C.J. Stam, E.C. van Straaten, E. Van Dellen,  P. Tewarie, G. Gong, A. Hillebrand, J. Meier and P. Van Mieghem,
``The relation between structural and functional connectivity patterns in complex brain networks",
{\it Int. J. Psychophysiol.} {\bf 103}, 149-160 (2016).

\bibitem{garces2016} P. Garc\'es, E. Pereda, J.A. Hern\'andez-Tamames, F. Del-Pozo, F. Maest\'u and J. Pineda-Pardo
``Multimodal description of whole brain connectivity: A comparison of resting state MEG, fMRI, and DWI",
{\it Human Brain Mapping} {\bf 37}, 20-34 (2016).

\bibitem{battiston2016} F. Battiston, V. Nicosia, M. Chavez and V. Latora,
``Multilayer motif analysis of brain networks",
https://arxiv.org/abs/1606.09115.

\bibitem{skardal2016d} P. S. Skardal, D. Taylor, J. Sun, and A. Arenas,
``Collective frequency variation in network synchronization and reverse PageRank'',
{\it Phys. Rev. E} {\bf 93}, 042314 (2016).

\bibitem{BenIsrael} A. Ben-Israel and T.N.E. Grenville,
{\it Generalized Inverses.} Springer, New York (1974).




\end{thebibliography}
\end{document}